\documentclass[twocolumn,showpacs,preprintnumbers,amsmath,amssymb]{revtex4}

\usepackage{graphicx}
\usepackage{dcolumn}
\usepackage{bm}

\begin{document}


\title{New experimental proposals for testing Dirac equation}
\author{Abel Camacho}
\email{acq@xanum.uam.mx}
\affiliation{Departamento de F\'{\i}sica \\
Universidad Aut\'onoma Metropolitana--Iztapalapa\\
Apartado Postal 55--534, C.P. 09340, M\'exico, D.F., M\'exico.}

\author{Alfredo Mac\'{\i}as}
\email{amac@xanum.uam.mx}
\affiliation{Departamento de F\'{\i}sica \\
Universidad Aut\'onoma Metropolitana--Iztapalapa\\
Apartado Postal 55--534, C.P. 09340, M\'exico, D.F., M\'exico.}

\date{\today}

\begin{abstract}
The advent of phenomenological quantum gravity has ushered us in
the search for experimental tests of the deviations from general
relativity predicted by quantum gravity or by string theories, and
as a by--product of this quest the possible modifications that
some field equations, for instance, the motion equation of
spin--1/2--particles, have already been considered. In the present
work a modified Dirac equation, whose extra term embraces a
second--order time derivative, is taken as mainstay, and three
different experimental proposals to detect it are put forward. The
novelty in these ideas is that two of them do not fall within the
extant approaches in this context, to wit, red--shift, atomic
interferometry, or Hughes--Drever type--like experiments.
\end{abstract}

\maketitle

One of the bedrocks beneath our present description of the
fundamental laws of physics is embodied by Lorentz symmetry. The
significance of this symmetry in the theo\-retical realm clearly
justifies the long--lasting inte\-rest in testing it
\cite{[1],[2], [3],[4]}. One of the profits in this context, the
one can be readily seen with a fleeting glimpse to the
corres\-ponding experimental constructions, is the fact that the
involved precisions have undergone a remarkable improvement.

The struggle in the quest for a quantum theory of gravi\-ty, and
the possibility of testing the different current approa\-ches
\cite{[5]} have rendered some predictions about the modified field
equations governing the motion of spin--1/2-- particles, induced
either by loop quantum gravity \cite{[6]}, or by string theory
\cite{[7]}.

Amid the gamut of predicted effects we may find the presence of
non--scalar mass terms, higher--order spatial derivatives, etc.,
\cite{[8]}. Nevertheless, a thorough analysis in this context
shall consider more general modifications to Dirac equation. For
instance, the emergence of higher--order spatial derivatives must
force us to mull over the appearance of higher--order time
derivatives as part of a physically relevant possibility. It is in
this last topic that the present work will delve. Forsooth, a
second--order time derivative term will be considered as a
primordial part of Dirac equation, and three new experimental
proposals, whose intention is the detection of this additional
contribution, will be put forward. Not only these ideas are
independent from each other, but also two of them do not fall
within the extant approaches in this context, to wit, red--shift,
atomic interferometry, or Hughes--Drever type--like experiments
\cite{[8]}.

The first idea addresses the dependence, upon the group velocity,
of the spreading of a wave packet. It will be shown that, in
principle, it is possible to detect higher--order time derivatives
monitoring the so--called spreading time of a wave packet.

The second proposal will take advantage of the fact that the
corresponding pro\-bability density displays a dependence, not
only, upon the added term, but also upon the sign of the electric
charge of the considered particle, a trait absent in the usual
theory.

Finally, in the last idea we will use the fact that Larmor
precession is, as will be shown later, a function of the extra
term, and in consequence the angular velocity of the expectation
values of the components of the spin allow us, in principle, to
test our modified Dirac equation.

In addition the feasibility of implementing in an experi\-mental
effort each one of the proposed models is also, briefly,
addressed.



As has been previously mentioned, our mainstay is the introduction
of a second--order time derivative in Dirac equation. To wit, from
square one we assume the following motion equation
\begin{equation}
i \hbar \frac{\partial}{\partial t} \psi = - i \hbar c
\mbox{\boldmath$\alpha$} \cdot \mbox{\boldmath$\nabla$} \psi +
\beta m c^2 \psi + \epsilon \frac{\hbar^2}{m c^2}
\frac{\partial^2}{\partial t^2}\psi\, . \label{DiracMemory}
\end{equation}
A factor $1/m c^2$ in the term containing the second--order time
derivative has been introduced, and the reason for this lies in
the convenience of having a dimensionless parameter $\epsilon$. It
is also readily seen that for $\epsilon = 0$, the introduced
equation reduces to the usual Dirac si\-tuation. Additionally, a
fleeting glimpse to (\ref{DiracMemory}) shows us that
Lorentz--covariance is violated.

This kind of modified Dirac equation has already been considered
\cite{[8]}, and also some experimental proposals for the detection
of the new contribution have been put forward. At this point it is
noteworthy to comment that all the aforementioned experiments fall
within the realm of interferometry, red--shift, or atomic
spectroscopy \cite{[8]}.

In the usual Dirac equation the non--relativistic limit is deduced
by splitting up the energy into two parts, namely, (i) the rest
energy, and (ii) additional contributions to the energy. This is
attained introducing
\begin{equation}
\psi = \tilde{\psi}\exp\Bigl(-\frac{i}{\hbar}mc^2t\Bigr)\, .
\end{equation}

The non--relativistic limit is obtained assuming that the rest
energy is much larger than any other kind of energy involved.
Proceeding as usual \cite{[9]}, which means that here

\begin{equation}
\tilde{\psi} = \left ({\begin{array}{c}
               \phi \\
               \chi \\
               \end{array}}
               \right)\, ,
\end{equation}
we arrive at the following expression
\begin{equation}
i[1- 2\epsilon(1 + \epsilon)]\hbar\frac{\partial \phi}{\partial t}
= -\frac{\hbar^2}{2m}\mbox{\boldmath$\nabla$}^2\phi
-\epsilon\frac{\hbar}{c}\lambda\frac{\partial^2 \phi}{\partial
t^2} + mc^2\epsilon^2(2 + \epsilon)\phi. \label{Nonrel}
\end{equation}
Here $\lambda$ denotes the Compton wavelength of the particle. The
presence of the last term in (\ref{Nonrel}) requires further
explanation. Indeed, it is readily seen that we do not know the
order of magnitude of $\epsilon$. In other words, even if
$\epsilon$ is very small, the term $mc^2\epsilon^2(2 + \epsilon)$
could have an order of magnitude similar to the remaining energies
present in (\ref{Nonrel}).


The introduction of spin is relevant, not only because it is a
fundamental physical trait, but also because one of the proposals
requires the interaction of a magnetic field with spin.
Accordingly, now we write down the genera\-lized Dirac equation,
considering its interaction with an electromagnetic field, and
afterwards, its corresponding Pauli equation will be derived.

The introduction of the coupling with an electromagnetic field is
achieved resorting to the minimal coupling procedure [8].
Therefore the resulting equation reads
\begin{eqnarray}
i \hbar \frac{\partial}{\partial t} \psi &=& - i \hbar c
\mbox{\boldmath$\alpha$} \cdot \Bigl(\mbox{\boldmath$\nabla$} -
\frac{iq}{\hbar c}\mbox{\boldmath$A$}\Bigr)\psi + \beta m c^2 \psi
\nonumber \\ &+& \epsilon
\frac{\lambda\hbar}{c}\Bigl(\frac{\partial}{\partial t} -
iq\Phi\Bigr)^2 \psi  + q\Phi\psi\, .  \label{DiracMemoryII}
\end{eqnarray}
In (\ref{DiracMemoryII}) we have introduced the vector potential,
$\mbox{\boldmath$A$}$, and the scalar one, $\Phi$. The
non--relativistic limit of this last expression renders the
generalized Pauli equation
\begin{eqnarray}
&&\hspace{-0.5cm}i[1- 2\epsilon(1 + \epsilon)]\hbar\frac{\partial
\phi}{\partial t} = \frac{\Bigl(-i\hbar\mbox{\boldmath$\nabla$}
-(q/c)\mbox{\boldmath$A$}\Bigr)^2}{2m}\phi + q\Phi\phi\nonumber
\\&-& \hspace{-0.5cm}+ \epsilon \frac{\lambda\hbar}{c}\Bigl(\frac{\partial}{\partial t} -
iq\Phi\Bigr)^2\phi + mc^2\epsilon^2(2 + \epsilon)\phi +
\frac{q}{mc}\mbox{\boldmath$S$}\cdot\mbox{\boldmath$B$}\phi \, .
\label{NonrelII}
\end{eqnarray}
Two new terms have been introduced in (\ref{NonrelII}), to wit,
the magnetic field, $\mbox{\boldmath$B$}$, and the spin operator,
$\mbox{\boldmath$S$}$, respectively.



Let us now consider a solution to (\ref{Nonrel}) in the form
\begin{equation}
\phi \sim
\exp\Bigl[i\Bigl(\mbox{\boldmath$k$}\cdot\mbox{\boldmath$r$} -
\omega t\Bigr)\Bigr]. \label{SolI}
\end{equation}
This Ansatz allows us to cast (\ref{Nonrel}) in the following form
\begin{equation}
[1- 2\epsilon(1 + \epsilon)]\hbar\omega = \frac{\hbar^2k^2}{2m} +
\epsilon\frac{\hbar}{c}\lambda\omega^2 + mc^2\epsilon^2(2 +
\epsilon). \label{NonrelIII}
\end{equation}
It is readily seen that this last expression defines $\omega$ as a
function of $k$. Indeed,
\begin{eqnarray}
& &\hspace{-0.5cm}\omega(k) = \frac{1}{2\epsilon\lambda}\Bigl\{[1-
2\epsilon(1 + \epsilon)]c \pm \nonumber
\\& &\hspace{-0.8cm} c\sqrt{[1- 2\epsilon(1 + \epsilon)]^2
-4\frac{\epsilon\lambda}{c\hbar}[mc^2\epsilon^2(2 + \epsilon) +
\frac{\hbar^2k^2}{2m}]}\Bigr\}\, . \label{Omega}
\end{eqnarray}

Quantum Mechanics \cite{[10]} teaches us that group and phase
velocity are defined by, $\nu_g = \frac{d\omega}{dk}$ and $\nu_p =
\frac{\omega}{k}$, respectively. Taking into account (\ref{Omega})
we obtain
\begin{equation}
\nu_g = \frac{\hbar k}{m}\Bigl\{[1- 2\epsilon(1 + \epsilon)]^2 -
4\frac{\epsilon\lambda}{c\hbar}[mc^2\epsilon^2(2 + \epsilon) +
\frac{\hbar^2k^2}{2m}]\Bigr\}^{-1/2}. \label{Group}
\end{equation}

An interesting point concerning the consequences of (\ref{Omega})
is cognate with the fact that it defines a cutoff in the permitted
wave number. Forsooth, the square--root, in (\ref{Omega}),
entails, in order to have real--valued frequency, the following
condition
\begin{equation}
k\leq
\sqrt{\frac{2m}{\hbar^2}}\Bigl\{\frac{c\hbar}{4\epsilon\lambda}[1
- 2\epsilon(1 + \epsilon)]^2 - mc^2\epsilon^2(2 +
\epsilon)\Bigr\}^{1/2}. \label{Cutoff}
\end{equation}

Assuming $\vert\epsilon\vert <<1$, the cutoff, in terms of the
momentum becomes, approximately
\begin{equation}
p \leq \frac{mc}{\sqrt{6\epsilon}}. \label{CutoffI}
\end{equation}
A condition always fulfilled within the non--relativistic realm.
Consider now a one--dimensional wave packet constructed as a
superposition of plane waves, in such a way that this packet is
sharply peaked around $k = k_0$, with a width given by $\Delta k$
\begin{equation}
\psi(x, t) =
\frac{1}{\sqrt{2\pi}}\int_{-\infty}^{\infty}A(k-k_0)\exp\Bigl\{ikx
- i\omega t\Bigr\}dk. \label{Wave}
\end{equation}
The condition upon the manner in which this wave packet has been
constructed implies that $A(k - k_0) \approx 0$ if $\vert k -
k_0\vert > \Delta k$. Expanding $kx - wt$ around $k = k_0$ allows
us to cast (\ref{Wave}) in the following form
\begin{eqnarray}
\psi(x, t) &=& \exp\Bigl\{ik_0x - i\omega(k_0) t\Bigr\}
\frac{1}{\sqrt{2\pi}}\int_{-\infty}^{\infty}A(q)\nonumber \\
&\times& \exp\Bigl\{iq\Bigl(x - [\nu_g -
q\frac{d^2\omega}{dk^2}_{\vert k_0}]t\Bigr)\Bigr\}dq.
\label{WaveI}
\end{eqnarray}
Here we have defined $q = k - k_0$. Since it has been assumed from
the very beginning that $A(k - k_0) \approx 0$ if $\vert k -
k_0\vert
>\Delta k$, then (\ref{WaveI}) will be dominated by values of $q$
in the range $[-\Delta k, \Delta k]$. Hence, we are allowed to put
forward the following relation
\begin{equation}
q\frac{d^2\omega}{dk^2}_{\vert k_0} = \pm\Delta\nu_g. \label{Conn}
\end{equation}

Knowing that the Fourier transform is dominated by those parts
satisfying the condition $x - \nu_gt \approx 0$ (as long as
$(\Delta k)^2\frac{d^2\omega}{dk^2}_{\vert k_0}t <<1$), then it is
reasonable to define the spreading time of the wave packet as
\begin{equation}
t_s = \Bigl[(\Delta k)^2\frac{d^2\omega}{dk^2}_{\vert
k_0}\Bigr]^{-1}. \label{Time}
\end{equation}
To first order in $\epsilon$ this spreading time reads
\begin{equation}
t_s = \frac{m}{\hbar(\Delta k)^2}\Bigl\{1 - 2\epsilon\Bigl[1 -
\frac{\lambda\hbar k_0^2}{2mc}\Bigr]\Bigr\}. \label{TimeI}
\end{equation}


Let us now hark back to (\ref{NonrelII}), with the initial
assumption of vanishing magnetic field, namely,
$\mbox{\boldmath$B$} = 0$. Proceeding in the usual manner [10] it
is possible to deduce a probability conservation law associated to
(\ref{NonrelII}). Indeed, under these circumstances
\begin{equation}
\frac{\partial\rho}{\partial t} +
\mbox{\boldmath$\nabla$}\cdot\mbox{\boldmath$J$} = 0, \label{Law}
\end{equation}
with
\begin{eqnarray}
\rho &=& \Bigl\{1 - \epsilon\frac{q\Phi\lambda}{c[1- 2\epsilon(1 +
\epsilon)]}\Bigr\}\phi\phi^{\ast} - i\epsilon\frac{\lambda}{c[1-
2\epsilon(1 + \epsilon)]}\nonumber \\
&\times&\Bigl\{\phi\frac{\partial\phi}{\partial t}^{\ast} -
\phi^{\ast}\frac{\partial\phi}{\partial t}\Bigr\}, \label{LawI}
\end{eqnarray}
and
\begin{eqnarray}
\mbox{\boldmath$J$} &=& i\frac{\hbar}{2m[1- 2\epsilon(1 +
\epsilon)]} \Bigl[\phi\mbox{\boldmath$\nabla$}\phi^{\ast} -
\phi^{\ast}\mbox{\boldmath$\nabla$} \phi\Bigr]\nonumber \\ &-&
q\frac{\lambda}{\hbar[1- 2\epsilon(1 +
\epsilon)]}\mbox{\boldmath$A$} \phi\phi^{\ast}. \label{LawII}
\end{eqnarray}
If $\epsilon =0$ is implemented, then everything reduces to the
usual conservation law [11]. The probability density not only
hinges upon first--order time derivatives, it also displays a
dependence on the charge of the involved particle. Both
characteristics are absent in the usual model \cite{[11]}.


Let us now analyze the case in which spin has to be considered,
and see if there is, in this context, enough leeway to pose an
experimental proposal that could detect the extra term. As shown
previously, the non--relativistic limit is embodied by
(\ref{NonrelII}). Henceforth it will be assumed that our involved
particle is at rest and that the magnetic field has non--vanishing
component only along the $z$--axis, i.e., $\mbox{\boldmath$B$} =
B_0\mbox{\boldmath$k$}$, where $B_0$ is a constant with dimensions
of magnetic field, and $\mbox{\boldmath$k$}$ denotes the unit
vector along the $z$--axis. Under these restrictions the dynamics
of the spin part of the system reads (here we have wri\-tten the
spin state ket as $\vert\chi > = \alpha\vert +> +\, \beta\vert
->$, where $S_z\vert \pm> = \pm\frac{\hbar}{2}\vert\pm>$)
\begin{equation}
i[1- 2\epsilon(1 + \epsilon)]\hbar\frac{d\alpha}{dt} =
-\epsilon\lambda\frac{\hbar}{c}\frac{d^2\alpha}{dt^2} +
\frac{q\hbar}{2mc}B_0\alpha, \label{LarmorII}
\end{equation}

\begin{equation}
i[1- 2\epsilon(1 + \epsilon)]\hbar\frac{d\beta}{dt} =
-\epsilon\lambda\frac{\hbar}{c}\frac{d^2\beta}{dt^2} -
\frac{q\hbar}{2mc}B_0\beta. \label{LarmorIII}
\end{equation}
It is readily seen that the solutions to these equations are (to
second order in $\epsilon$) given by
\begin{eqnarray}
\vert\chi> &=&
\cos\Bigl(\frac{\theta}{2}\Bigr)\exp\Bigl\{-i\frac{qB_0}{2mc}[1 +
2\epsilon(1 + \epsilon)]t\Bigr\}\vert +> \nonumber \\
&+&
\sin\Bigl(\frac{\theta}{2}\Bigr)\exp\Bigl\{i\frac{qB_0}{2mc}[1+
2\epsilon(1 + \epsilon)]t\Bigr\}\vert ->. \label{LarmorIV}
\end{eqnarray}
In the last expression $\theta$ depends upon the initial
conditions of the spin state ket. The condition $\epsilon = 0$
renders the usual situation \cite{[11]}. If now the expectation
value for $S_x$ is evaluated we find that
\begin{equation}
<S_x>_{\chi} =
\frac{\hbar}{2}\sin\Bigl(\theta\Bigr)\cos\Bigl\{\frac{qB_0}{mc}[1
+ 2\epsilon(1 + \epsilon)]t\Bigr\}. \label{Sx}
\end{equation}
From (\ref{Sx}) the frequency of this modified Larmor prece\-ssion
is easily read off
\begin{equation}
\omega = \frac{\vert q\vert B_0}{mc}[1 + 2\epsilon(1 + \epsilon)].
\label{Fre}
\end{equation}


Let us now address the feasibility of the aforementioned
experimental proposals. Firstly, the possibility of resorting to
the spreading time of a wave packet in order to detect an extra
term, like the one encompassed by  (\ref{DiracMemory}), is cognate
with the fact that the experimental resolution, $\Delta t$, has to
be smaller than the difference between the spreading times in our
proposal, (\ref{TimeI}), and the spreading time in the usual
model, henceforth denoted by $\tilde{t}_s$, where
$\tilde{t}_s = \frac{m}{\hbar(\Delta k)^2}.$ 
In other words, it will be po\-ssible to detect, within the realm
of the first proposal, an extra term like the one here considered
if
\begin{equation}
\Delta t < \frac{2m}{\hbar(\Delta k)^2}\Bigl\{1 -
\frac{\lambda\hbar k_0^2}{2mc}\Bigr\}\vert\epsilon\vert .
\label{TimeIII}
\end{equation}
This last expression may be used to set a bound, in the case of a
null experiment, to the magnitude of $\epsilon$. Forsooth, if an
experiment renders no evidence of this kind of extra term, then it
means that
\begin{equation}
\vert\epsilon\vert < \frac{\hbar(\Delta k)^2}{2m}\Bigl\{1 -
\frac{\lambda\hbar k_0^2}{2mc}\Bigr\}^{-1}\Delta t .
\label{TimeIV}
\end{equation}
Usually \cite{[8]} the tests (which employ as probes quantum
systems) of the postulates behind general relativity are divided
into three different types: (i) Hughes--Drever type--like ideas,
(ii) red--shift experiments, and (iii) interferometry.  The latter
is sensitive to the center of mass motion of quantum systems,
whereas the former probes the energy of bound states.  Clearly,
the spreading time of a wave packet has no classical analogue, and
in consequence the first proposal is a new test of Lorentz
covariance, the one is not encompassed by neither of the three
aforementioned ideas.

{\bf Beware, new material inserted here.}

Let us now put forward a particular experimental setup designed to
detect, within the context of spreading time, the magnitude of
$\epsilon$.

Consider a particle at rest, whose wave function embodies a linear
superposition of plane waves, in such a way that its initial form
is gaussian (the maximum of the norm of the wave function will be
at the origin of the coordinate system). Two screens will be
located at two different points, such that they initially lie
outside the root--mean square--deviation in the corresponding
space variable. In other words, if the positions of the
aforementioned screens are denoted by $ 0< S_1 < S_2$ and $\Delta
x(t=0)$ is the root--mean--square--deviation at time $t =0$, then
$\Delta x(t=0)< S_1$. As time goes by the packet spreads, and in
consequence a time will come, say $t_1$, in which $\Delta x(t=t_1)
= S_1$. Screen 1, at this moment, emits a photon. The same
situation will be associated to the se\-cond screen, to wit, at
time $t_2$, the root--mean--square--deviation fulfills $\Delta
x(t=t_2) = S_2$. The time interval between these two photons will
be related to the spreading velocity of the packet, and since we
know the distance between the two screens, $S_2 - S_1$, then the
knowledge of these two factors would allow us to set a bound to
the $\epsilon$ parameter, the one appears in the spreading time,
and in consequence in the spreading velocity. The possi\-bility of
measuring time intervals down to 50 fs is already within the
technological developments. The experimental method is founded
upon a fourth--order interference technique between two photons,
and it permits the pre\-sence of an accuracy of $1$ fs
\cite{[12]}.

Therefore, we may reduce the measuring of the spreading velocity
of the wave packet to the measuring of the time interval between
two photons, which is a case that nowadays can be done with a very
good precision \cite{[12]}.

{\bf Beware, new material finishes here.}

 The possibility of
employing the probability density to detect the extra term is
related to the fact that, as  (\ref{LawI}) clearly displays, the
probability density hinges upon the charge of the corresponding
particle, whereas in the usual theory it does not. Therefore, if
we perform the change $q\to -q$, then the aforementioned
expression leads us to conclude that there must be a change in the
probability density. This change in the probability density
associated to the modification of the charge of the involved
particle is not present in the usual situation, and defines a
trait that could, in principle lead to the detection of the new
term.

For the sake of clarity let us assume that in our experiment we
prepare the system such that $\phi\frac{\partial\phi}{\partial
t}^{\ast} - \phi^{\ast}\frac{\partial\phi}{\partial t} =0$, at $t
=0$. It would be possible to detect the extra term if, here
$\Delta\rho$ denotes the  experimental resolution in the measuring
process of the probability density
\begin{equation}
\Delta\rho < \vert q\epsilon\vert\frac{\Phi\lambda}{c[1-
2\epsilon(1 + \epsilon)]}\phi\phi^{\ast}. \label{LawIV}
\end{equation}
Additionally, $\rho$, in the present model, has a
time--dependence, embodied in the last term depicted in
(\ref{LawI}), the one does not emerge in the usual theory. The
concept of probability density has not been used to detect any
kind of violation to Lorentz covariance, and  a fleeting glimpse
to the current proposals \cite{[8]} readily shows us that the
second proposal does not fall within the usual experimental ideas.

{\bf Beware, new material inserted here.}

Let us now introduce the possibility of detecting $\epsilon$ with
the interaction embodied in (\ref{NonrelII}). In order to do this
we hark back to this aforementioned expression and take a very
particular case, namely, we choose $\mbox{\boldmath$A$} =0$.
Henceforth, the dynamics does not embrace the spin of our
particle, the one to first order in $\epsilon$ has the following
face

\begin{eqnarray}
&&\hspace{-0.5cm}i[1- 2\epsilon(1 + \epsilon)]\hbar\frac{\partial
\phi}{\partial t} =
\frac{\Bigl(-i\hbar\mbox{\boldmath$\nabla$}\Bigr)^2}{2m}\phi +
q\Phi\phi\nonumber
\\&-& \hspace{-0.5cm}+ \epsilon \frac{\lambda\hbar}{c}\Bigl(\frac{\partial}{\partial t} -
iq\Phi\Bigr)^2\phi. \label{ScattI}
\end{eqnarray}

A fleeting glimpse to (\ref{Nonrel}) (keeping only terms of first
order in $\epsilon$) clearly shows us that we may interpret the
presence of the term $[1- 2\epsilon(1 + \epsilon)]$ as a
redefinition of the inertial mass parameter as follows

\begin{eqnarray}
&&\hspace{-0.5cm}\tilde{m} = m[1- 2\epsilon(1 + \epsilon)].
\label{Mass}
\end{eqnarray}

It is a very know fact that scattering of particles has been a
useful tool in physics. Indeed, a lot of the most important
discoveries in physics have been achieved with the help of this
method \cite{[13]}. The idea in this part of the work is to take
advantage of the experience within this context, and try to put
forward a physical quantity that could be measured, and which
should render information about $\epsilon$. In this spirit, we may
confront (\ref{ScattI}) against experimental evidence noting that
in a scattering expe\-riment, in the low--energy limit, the Born
approximation entails the presence of the inertial mass parameter
\cite{[14]} for the scattering amplitude

\begin{eqnarray}
&&\hspace{-0.5cm}f(\theta, \Phi) = -\frac{m}{2\pi\hbar^2}\int
V(\vec{r})d^3\vec{r}. \label{ScattamI}
\end{eqnarray}

The comment regarding the redefinition of the inertial mass
parameter leads us to conclude that in the gene\-ralized
Schr\"odinger equation the corresponding scattering amplitude (to
first order in $\epsilon$) becomes (for spherical symmetry)

\begin{eqnarray}
&&\hspace{-0.5cm}f(\theta) = -\frac{2m[1 -
2\epsilon]}{\hbar^2\kappa}\int_0^{\infty} rV(r)\sin(\kappa r)dr.
\label{ScattamII}
\end{eqnarray}

In this last equation we have introduced an additional parameter,
to wit, $\kappa = 2k\sin(\theta/2)$. The connection with the
experiment is deduced immediately recalling that the differential
cross section $\frac{d\sigma}{d\Omega}$ is given by

\begin{eqnarray}
&&\hspace{-0.5cm}\frac{d\sigma}{d\Omega} = \vert f(\theta)\vert^2.
\label{CrossI}
\end{eqnarray}

To first order in $\epsilon$ we have that (here
$\frac{d\sigma}{d\Omega}_U$ denotes the differential cross section
in the usual model)

\begin{eqnarray}
&&\hspace{-0.5cm}\frac{d\sigma}{d\Omega} = [1 -
4\epsilon]\frac{d\sigma}{d\Omega}_U. \label{CrossII}
\end{eqnarray}

The proposed experiment could be carried out using electrons, in
the long--wavelength limit, which should impinge upon a
spherically symmetric scattering potential. This kind of
experiments, as has been mentioned above, comprise already a good
deal of experience, and in consequence this proposal lies within
the present technological possibilities. The current precision
associated to the detection of the number of particles scattered
off would then define a bound to the magnitude of $\epsilon$.

 {\bf Beware, new material finishes here.}

 Finally, a modified
Larmor precession entails an additional manner to detect
$\epsilon$. Looking at (\ref{Fre}) it is easily seen that (in the
usual case the Larmor frequence reads $\tilde{\omega} =
\frac{qB_0}{mc}$ \cite{[11]}) within this idea we need a time
resolution, $\Delta t$, fulfilling
\begin{equation}
\Delta t < 2\epsilon\frac{mc}{\vert q \vert B_0}\Bigl(1 -
\epsilon\Bigr). \label{FreI}
\end{equation}

{\bf Beware, new material inserted here.}

 The feasibility of this last idea is cognate with the cu\-rrent
 technological precision related to the measurement of the so called Bohr
 frequency. Indeed, Larmor expre\-ssion appears for the frequency of an atom, immersed in a
 uniform magnetic field, related to the energy eigenva\-lues of the Hamiltonian which describes
 the spin evolution \cite{[15]}. Hence, the idea at this point is to exploit this fact, and
in consequence in this last part of the
 present work the proposed experiment consists
 in the measurement of the energy difference of this kind of atoms,
 for instance, a silver atom, which is a system already studied within this realm.
 Indeed, denoting by $E_+$ and $E_-$
 the two corresponding le\-vels it is readily seen that the present
 idea leads us to look for deviations in the silver atom for the
 aforementioned energy difference, which in our case is tantamount
 to ($\Delta E_U = \hbar\tilde{\omega}/(2\pi)$ denotes the energy difference in the usual
 theory \cite{[16]})

\begin{equation}
\Delta E= \Delta E_U[1 + 2\epsilon]. \label{EndiffI}
\end{equation}

Though the extant literature already comprises results that
evaluate the shift in the energy levels, for instance, of a
hydrogen atom, our proposal involves the effects of the new term
upon spin, a fact that seems to require further analysis. The
question regarding the feasibility of the present proposal poses
no difficulty, since this kind of experiments in spectroscopy have
been already carried out \cite{[17]}).

 {\bf Beware, new material finishes here.}

Summing up, quantum gravity and string theories entail possible
modifications to some field equations, and in this realm our
initial premise has been a modified Dirac equation, which embraces
a second--order time derivative. The main idea in the present work
delves with the detection of the aforementioned new term putting
forward three new experimental proposals. At this point it is
noteworthy to mention that two of them do not fall within the
usual cases, either atomic interferometry, red shift, or
Hughes--Drever type--like experiments.  Finally, it is also
feasible to detect this kind of modifications to Dirac equation
looking at the changes that emerge in the context of Berry's
phase. The results in this issue will be published elsewhere.

This research was supported by CONACYT Grant 42191--F. A.C. would
like to thank A.A. Cuevas--Sosa for useful discussions and
literature hints.

\end{document}